# Control Performance Analysis of Power Steering System Electromechanical Dynamics

**Prerit Pramod**, *Senior Member*, IEEE

Control Systems Engineering, MicroVision, Inc.
*Email*: preritpramod89@gmail.com; preritp@umich.edu; prerit_pramod@microvsion.com

**Abstract**: Modern power steering systems employ an electric motor drive system to provide torque assistance to the driver. The closed-loop mechanical system dynamics that impact stability, performance and steering feel are significantly impacted by the electrical dynamics of the actuator depending on the structure and tuning of the motor torque controller. This paper presents an integrated approach to the analysis of this electromechanical dynamic control interaction through mathematical modeling which is confirmed with simulations.

## Introduction

Electric power steering (EPS) systems are virtually ubiquitous in the automotive industry, particularly the passenger vehicle segment, owing to the myriad of advantages they offer in terms of fuel savings, tunability and performance flexibility and inherent intelligence in the form of fault detection and correction [1]–[5].

Rapidly changing vehicle safety and performance requirements have mandated the need for electric motor drives that provide rapid dynamics, variable bandwidth and fault tolerance [6]–[9]. Permanent magnet synchronous machines (PMSMs) are most widely employed in EPS systems owing to the multifarious advantages they offer including, but not limited, high power density, low inertia and relatively minimal torque ripple [10]–[12]. The torque control of PMSMs used in EPS systems is performed indirectly via current control which may be implemented in multifarious forms. Feedforward or open loop control architectures employ an inverse mathematical model of the machine to determine the voltage commands to be applied for given current commands [13]–[20] . Alternatively, feedback control architectures involve the closed loop regulation of measured currents [21]–[30]. Depending on the selected control architecture and tuning of control parameters, the motor control system may significantly impact the overall stability and performance of EPS systems, which is typically ignored and results in sub-optimal control design and tuning of the overall system.

This paper presents a detailed mathematical treatment of overall EPS electromechanical dynamics through modeling of the interaction of different EMD torque control architectures with the mechanical system which has not been presented previously in literature.

## Electric Power Steering System Models

A typical EPS control architecture is shown in Fig. 1. The driver torque $T_d$ is sensed via a handwheel torque sensor signal $\hat{T}_h$ and is used to compute the motor (assistance) torque command $T_m^*$ along with other dynamic compensation. The EMD consisting of a PMSM then converts the torque command to current commands $I_{dq}^*$ in the so-called synchronous frame and commands voltages $V_{dq}^*$ which are applied to the motor through the gate driver and inverter through either a feedforward or feedback current control structure.

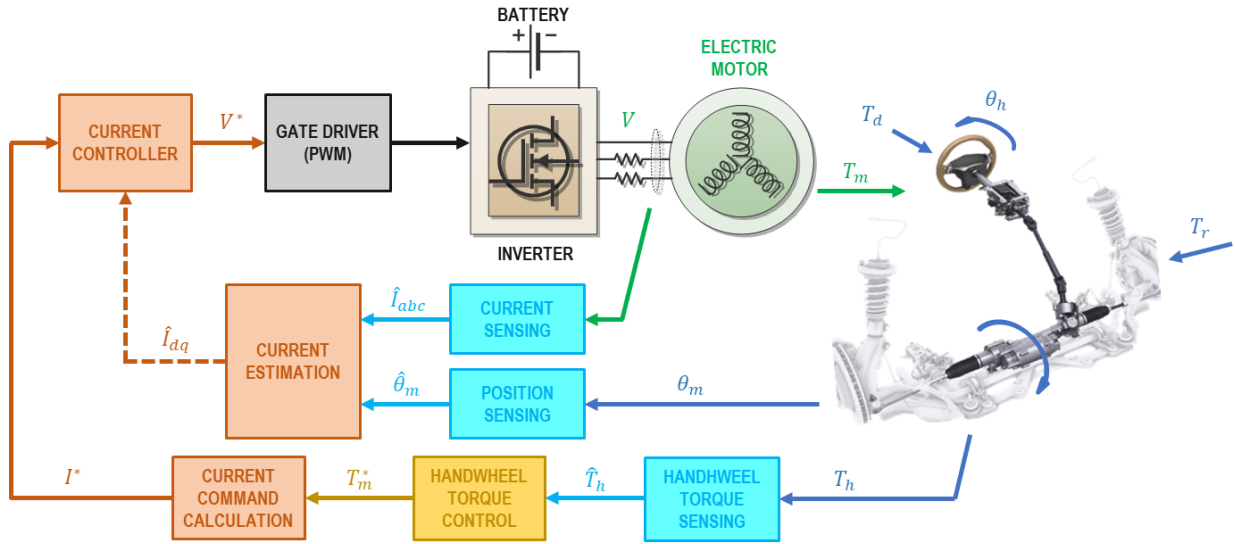

**Fig. 1**. Electric power steering control system.

The mechanical system may be modeled as a 2-mass model (2MM) wherein the handwheel and assist mechanism (plus the motor) are separated by the torque bar used as the handwheel torque sensor. The transfer functions relating the sensed handwheel torque to the external inputs of driver torque $T_d$ and rack force $T_r$ (translated to rotating coordinates) and the control signal motor torque $T_m$ are given in (1).

$$\begin{aligned}
T_h &= (N.M_t)T_m + (M_r)T_r + (M_d)T_d \\
&= \left(K_h \frac{K_h - M_h}{D}\right)T_m + \left(K_h \frac{K_h - M_h}{D}\right)T_r + \left(K_h \frac{M_m - K_h}{D}\right)T_d
\end{aligned} \quad (1)$$

$$M_h = J_h s^2 + b_h s \qquad M_m = J_m s^2 + b_m s + K_l \qquad D = M_h M_m - K_h^2$$

where $N$ is the gear ratio, $J$, $b$ and $K$ denote inertia, damping and compliance respectively, and subscripts $h$ and $m$ represents the handwheel and motor (including assist mechanism) respectively, while $K_l$ is the linear tire model spring constant. The conventional synchronous frame motor model is non-linear since it contains voltage terms involving a product of velocity and currents and is linearized as perturbations represented by $\Delta$ about a set-point ($V_{dq}^0$, $I_{dq}^0$, $\omega_m^0$) for the present analysis as given in (2).

$$T_m = q.p\lambda_m\left(I_q^0 + \Delta I_q\right)$$

$$V_{dq}^0 + \Delta V_{dq} = P^{-1}\left(I_{dq}^0 + \Delta I_{dq}\right) + E(\omega_m^0 + \Delta\omega_m)$$

$$\begin{bmatrix} V_d^0 + \Delta V_d \\ V_q^0 + \Delta V_q \end{bmatrix} = \begin{bmatrix} L_d s + R & p\omega_m^0 L_q \\ -p\omega_m^0 L_d & L_q s + R \end{bmatrix}\begin{bmatrix} I_d^0 + \Delta I_d \\ I_q^0 + \Delta I_q \end{bmatrix} + \begin{bmatrix} pL_q I_q^0 \\ -pL_d I_d^0 + p\lambda_m \end{bmatrix}(\omega_m^0 + \Delta\omega_m) \quad (2)$$

where $V$ and $I$ are voltage and current respectively, $\omega_m$ is motor velocity, $p$ is magnetic pole pairs, $\lambda_m$ is the permanent magnet flux linkage, $L$ represents inductance, $R$ is the combined motor and inverter resistance, $q$ is a constant equal to 1.5, and the subscripts $d$ and $q$ represent the direct and quadrature axis in the synchronous frame. Inverter and current estimation dynamics are modeled as transport lags of delay $\tau_c$ and $\tau_p$ as given in (3).

$$\hat{I}_{dq} = B I_{dq} \qquad V_{dq} = X V_{dq}^*$$

$$\begin{bmatrix} \hat{I}_d \\ \hat{I}_q \end{bmatrix} = \begin{bmatrix} e^{-\tau_c s} & 0 \\ 0 & e^{-\tau_c s} \end{bmatrix}\begin{bmatrix} I_d \\ I_q \end{bmatrix} \qquad \begin{bmatrix} V_d \\ V_q \end{bmatrix} = \begin{bmatrix} e^{-\tau_p s} & 0 \\ 0 & e^{-\tau_p s} \end{bmatrix}\begin{bmatrix} V_d^* \\ V_q^* \end{bmatrix} \quad (3)$$

## Electric Motor Drive Control Dynamics

Motor torque control may be implemented as either a feedforward, i.e., open loop (model based) or a feedback current control (utilizing measured current feedback) system. A general expression for overall electrical dynamics of the EMD may be written as a torque tracking transfer function $A_t$ and disturbance rejection transfer function $A_d$ as given in (4).

$$T_m = A_t T_m^* + A_\omega \omega_m \quad (4)$$

Under normal operating conditions, i.e., at low velocities, the motor current commands are computed from the torque command using the maximum torque per ampere (MTPA) technique as given in (5).

$$I_q^* = \frac{1}{qp\hat{\lambda}_m}T_m^*$$

$$I_d^* = 0 \quad (5)$$

### Feedforward Current Control

This open-loop technique involves calculation of the voltage commands using an inverse motor model involving estimated parameters as shown in (6).

$$V_{dq}^* = C_t^f I_{dq}^* + C_\omega^f \hat{\omega}_m$$

$$\begin{bmatrix} V_d^* \\ V_q^* \end{bmatrix} = \begin{bmatrix} \hat{L}_d \hat{s} + \hat{R} & p\hat{\omega}_m \hat{L}_q \\ -p\hat{\omega}_m \hat{L}_d & \hat{L}_q \hat{s} + \hat{R} \end{bmatrix} \begin{bmatrix} I_d^* \\ I_q^* \end{bmatrix} + \begin{bmatrix} 0 \\ p\hat{\lambda}_m \end{bmatrix} \hat{\omega}_m \tag{6}$$

The velocity estimate is modeled as a transfer function $H_\omega$ incorporating position sensing and velocity estimation dynamics as a derivative with low pass filtering as given in (7).

$$\hat{\omega}_m = H_\omega \omega_m = \left( \frac{s}{\tau_\omega s + 1} \right) \omega_m \tag{7}$$

Combining (2) through (7) at zero nominal velocity ($\omega_m^0 = 0$) results in the overall feedforward motor torque control transfer function shown in (8).

$$T_m = A_t^f T_m^* + A_\omega^f \omega_m$$

$$= X \frac{\lambda_m}{\hat{\lambda}_m} \left( \frac{\hat{L}_q s + \hat{R}}{L_q s + R} \right) T_m^* + qp^2 \lambda_m \left( \frac{X H_\omega \hat{\lambda}_m - \lambda_m}{L_q s + R} \right) \omega_m \tag{8}$$

### Feedback Current Control

A current regulator such as a PI controller is used to minimize the error between the commanded currents and the measured current feedback as given in (9).

$$V_{dq}^* = C_t^b (I_{dq}^* - \hat{I}_{dq}) + C_\omega^b \hat{\omega}_m$$

$$\begin{bmatrix} V_d^* \\ V_q^* \end{bmatrix} = \begin{bmatrix} K_{pd} + \frac{K_{id}}{\hat{s}} & 0 \\ 0 & K_{pq} + \frac{K_{iq}}{\hat{s}} \end{bmatrix} \begin{bmatrix} I_d^* - \hat{I}_d \\ I_q^* - \hat{I}_q \end{bmatrix} + \begin{bmatrix} 0 \\ p\hat{\lambda}_m \end{bmatrix} \hat{\omega}_m \tag{9}$$

Combining (2) through (5) with (7) and (9) at zero nominal velocity ($\omega_m^0 = 0$) results in the overall feedback motor torque control transfer function shown in (10).

$$T_m = A_t^b T_m^* + A_\omega^b \omega_m$$

$$= X \frac{\lambda_m}{\hat{\lambda}_m} \left( \frac{\hat{K}_{pq} \hat{s} + \hat{K}_{iq}}{L_q s \hat{s} + R\hat{s} + XBK_{pq}\hat{s} + XBK_{iq}} \right) T_m^* + qp^2 \lambda_m \left( \frac{X H_\omega \hat{\lambda}_m - \lambda_m}{L_q s \hat{s} + R\hat{s} + XBK_{pq}\hat{s} + XBK_{iq}} \right) \omega_m \tag{10}$$

## Steering Control Closed-loop Dynamical Analysis

The stability as well as performance of the overall EPS control system may be studied by assessing the effective open-loop transfer function (E-OLTF) relating $T_m^*$ to $\hat{T}_h \approx T_h$ obtained by combining (1) and (4) as given in (11).

$$T_h = Z_t T_m^* + Z_r T_r + Z_d T_d$$
$$= \left(K_h \frac{K_h - M_h}{D - sA_\omega M_h} A_t\right) T_m^* + \left(K_h \frac{K_h - M_h}{D - sA_\omega M_h}\right) T_r + \left(K_h \frac{M_m - K_h - sA_\omega}{D - sA_\omega M_h}\right) T_d \quad (11)$$

where $A_t$ and $A_\omega$ may be replaced by expressions from (8) and (10) to obtain the specific E-OLTFs for feedforward and feedback control respectively. The motor torque control transfer function in (12) including mechanical dynamics is derived by combining (1) and (11). The expansion and simplification of (11) and (12) for various cases is left for the full paper.

$$T_m = W_t T_m^* + W_r T_r + W_d T_d$$
$$= \left(\frac{D}{D - sA_\omega M_h} A_t\right) T_m^* + \left(\frac{sA_\omega M_h}{D - sA_\omega M_h}\right) T_r + \left(K_h \frac{sA_\omega}{D - sA_\omega M_h}\right) T_d \quad (12)$$

Simulation results depicting motor torque control frequency response modification due to EPS mechanical dynamics and the corresponding E-OLTFs are shown in Fig. 2.

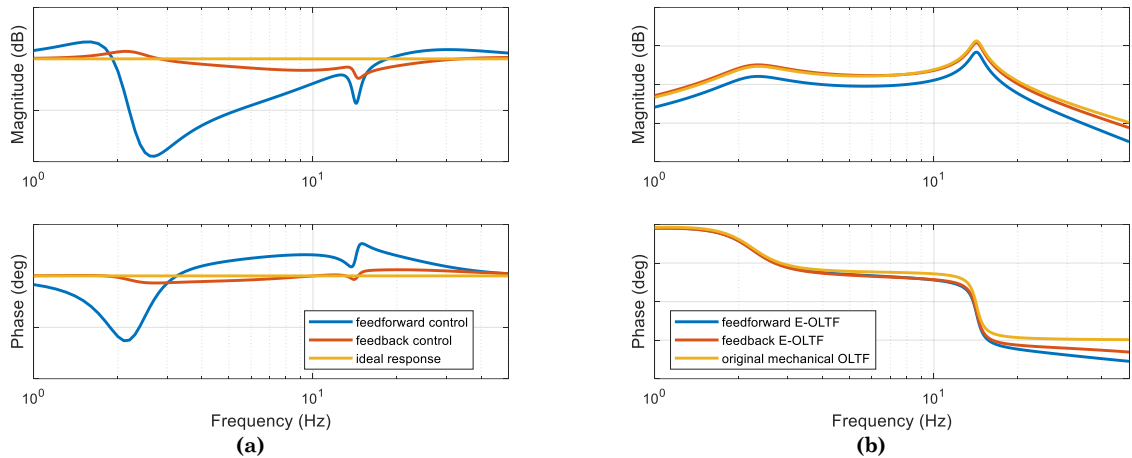

**Fig. 2** Simulation results illustrating frequency response plots of (a) motor torque control scaling due to mechanical dynamics (b) steering actual and effective open-loop system.

The motor control scaling responses shown in Fig. 2(a) illustrate the comparative behavior of the EMD with mechanical dynamics relative to electrical-only (ideal) control dynamics. The impact on EPS dynamics is clear from Fig. 2(b) which depicts a reduction in both the gain and phase (stability) margins when feedforward torque control rather than feedback control is employed, due to its inferior disturbance rejection characteristics.

## Conclusions

A comprehensive mathematical modeling capturing the overall closed-loop electromechanical dynamics of EPS systems considering the influence of EMD dynamics is presented. The novel model is crucial for properly characterizing, stabilizing and tuning EPS systems.

# References

[1] Xin Li, Xue-Ping Zhao, Jie Chen. "Controller Design for Electric Power Steering System Using T-S Fuzzy Model Approach." *International Journal of Automation and Computing* 6 (2): 198–203, 2009.

[2] Xiang Chen, Tiebao Yang, Xiaoqun Chen, Kemin Zhou. 2008. "A Generic Model-Based Advanced Control of Electric Power-Assisted Steering Systems." *IEEE Transactions on Control Systems Technology* 16 (6): 1289–1300, 2008.

[3] Manu Parmar, John Y. Hung. "A Sensorless Optimal Control System for an Automotive Electric Power Assist Steering System." *IEEE Transactions on Industrial Electronics* 51 (2): 290–98, 2004.

[4] Aly Badawy, Jeff Zuraski, Farhad Bolourchi, Ashok Chandy. "Modeling and Analysis of an Electric Power Steering System." In. Society of Automotive Engineers, 1999. doi:10.4271/1999-01-0399.

[5] Nilay Kant, Raunav Chitkara, Prerit Pramod. "Modeling Rack Force for Steering Maneuvers in a Stationary Vehicle". *SAE Technical Paper*, No. 2021-01-1287, 2021.

[6] Prerit Pramod, Priyanka Mendon, Chethan Narayanaswamy, Fischer Klein. "Impact of Electric Motor Drive Dynamics on Performance and Stability of Electric Power Steering Systems". *SAE Technical Paper*, No. 2021-01-0932, 2021.

[7] Prerit Pramod, Rakesh Mitra, Rangarajan Ramanujam. "Current sensor fault mitigation for steering systems with permanent magnet DC drives." *U.S. Patent 10,822,024*, issued November 3, 2020.

[8] Prerit Pramod, Kai Zheng, Mariam Swetha George, Tejas M. Varunjikar. "Cascaded position control architecture for steering systems." *U.S. Patent 11,203,379*, issued December 21, 2021.

[9] Prerit Pramod, Michael A. Eickholt, David P. Holm, Rakesh Mitra. "Fault tolerant control of rear steer vehicles." *U.S. Patent Application 16/709,162*, filed November 12, 2020.

[10] Prerit Pramod. 2023. "Circulating Currents in Delta Wound Permanent Magnet Synchronous Machines." *OSF Preprints*. June 21. doi:10.31219/osf.io/awqkr.

[11] Alejandro J. Pina, Prerit Pramod, Rakib Islam, Rakesh Mitra, Longya Xu. "Modeling and experimental verification of torque transients in interior permanent magnet synchronous motors by including harmonics in d-and q-axes flux linkages." In *2015 IEEE International Electric Machines & Drives Conference (IEMDC)*, pp. 398-404. IEEE, 2015.

[12] Alejandro J. Piña, Prerit Pramod, Rakib Islam, Rakesh Mitra, Longya Xu. "Extended model of interior permanent magnet synchronous motors to include harmonics in d-and q-axes flux linkages." In *2015 IEEE Energy Conversion Congress and Exposition (ECCE)*, pp. 1864-1871. IEEE, 2015.

[13] Prerit Pramod, Zhe Zhang, Rakesh Mitra, Subhra Paul, Rakib Islam, Julie Kleinau. "Impact of parameter estimation errors on feedforward current control of permanent magnet synchronous motors." In *2016 IEEE Transportation Electrification Conference and Expo (ITEC)*, pp. 1-5. IEEE, 2016.

[14] Prerit Pramod, Zhe Zhang, Krishna MPK Namburi, Rakesh Mitra, Subhra Paul, Rakib Islam. "Effects of position sensing dynamics on feedforward current control of permanent magnet synchronous machines." In *2017 IEEE International Electric Machines and Drives Conference (IEMDC)*, pp. 1-7. IEEE, 2017.

[15] Prerit Pramod, Rakesh Mitra, Krishna MPK Namburi, Aparna Saha, Infane Lowe. "Resistance Imbalance in Feedforward Current Controlled Permanent Magnet Synchronous Motor Drives." In *2019 IEEE Transportation Electrification Conference and Expo (ITEC)*, pp. 1-5. IEEE, 2019.

[16] Julie A. Kleinau, Prerit Pramod, Dennis B. Skellenger, Selva Kumar Sengottaiyan. "Motor control current sensor loss of assist mitigation for electric power steering." *U.S. Patent 9,809,247*, issued November 7, 2017.

[17] Prerit Pramod, Rakesh Mitra. "Feedforward control of permanent magnet DC motors." *U.S. Patent 10,404,197*, issued September 3, 2019.

[18] Prerit Pramod, Varsha Govindu, Zhe Zhang, Krishna Mohan Pavan Kumar Namburi. "Feedforward control of permanent magnet synchronous motor drive under current sensing failure." *U.S. Patent 10,717,463*, issued July 21, 2020.

[19] Prerit Pramod, Infane Lowe, Krishna MPK Namburi, Varsha Govindu. "Torque ripple compensation with feedforward control in motor control systems." *U.S. Patent 10,333,445*, issued June 25, 2019.

[20] Prerit Pramod. "Feedforward current control for dual wound synchronous motor drives." U.S. Patent 11,736,048, issued August 22, 2023.

[21] Prerit Pramod. "Synchronous frame current estimation inaccuracies in permanent magnet synchronous motor drives." In *2020 IEEE Energy Conversion Congress and Exposition (ECCE)*, pp. 2379-2386. IEEE, 2020.

[22] Prerit Pramod, Krishna MPK Namburi. "Closed-loop current control of synchronous motor drives with position sensing harmonics." In *2019 IEEE Energy Conversion Congress and Exposition (ECCE)*, pp. 6147-6154. IEEE, 2019.

[23] Prerit Pramod, Zhe Zhang, Krishna MPK Namburi, Rakesh Mitra, Darren Qu. "Effects of position sensing dynamics on feedback current control of permanent magnet synchronous machines." In *2018 IEEE Energy Conversion Congress and Exposition (ECCE)*, pp. 3436-3441. IEEE, 2018.

[24] Prerit Pramod. "Feedback compensation of parameter imbalance induced current harmonics in synchronous motor drives." *U.S. Patent 11,349,416*, issued May 31, 2022.

[25] Prerit Pramod, Shrenik P. Shah, Julie A. Kleinau, Michael K. Hales. "Decoupling current control utilizing direct plant modification in electric power steering system." *U.S. Patent 10,003,285*, issued June 19, 2018.

[26] Prerit Pramod, Julie A. Kleinau. "Current mode control utilizing plant inversion decoupling in electric power steering systems." *U.S. Patent 11,091,193*, issued August 17, 2021.

[27] Prerit Pramod, Julie A. Kleinau. "Motor control anti-windup and voltage saturation design for electric power steering." *U.S. Patent 10,103,667*, issued October 16, 2018.

[28] Prerit Pramod, Varsha Govindu, Rakesh Mitra, and Nithil Babu Nalakath. "Current regulators for permanent magnet DC machines." *U.S. Patent 10,640,143*, issued May 5, 2020.

[29] Prerit Pramod. "Current regulators for dual wound synchronous motor drives." *U.S. Patent Application 17/563,908*, filed June 29, 2023.

[30] Siddharth Mehta, Prerit Pramod, Iqbal Husain. "Analysis of dynamic current control techniques for switched reluctance motor drives for high performance applications." In *2019 IEEE Transportation Electrification Conference and Expo (ITEC)*, pp. 1-7. IEEE, 2019.